\documentclass{eas}
\usepackage{graphicx}
\usepackage{mathrsfs}
\usepackage{amssymb}
\usepackage{amsmath}
\usepackage{graphicx}% Include figure files
\usepackage{bm}% bold math
\begin{document}

%%-----------------------------
%%      the top matter
%%-----------------------------
\title{Track reconstruction with MIMAC} 
%
%\runningtitle{Phenomenology of Directional Detection of Dark Matter}
%
\author{J. Billard}\address{Laboratoire de Physique Subatomique et de Cosmologie, Universit\'e Joseph Fourier Grenoble 1,
  CNRS/IN2P3, Institut Polytechnique de Grenoble, Grenoble, France}
\author{F. Mayet}\sameaddress{1}
\author{D. Santos}\sameaddress{1}
%\author{...}\address{...}
%
%
\begin{abstract}
Directional detection of Dark Matter is a promising search strategy.
However, to perform such kind of detection, the recoiling tracks have to be accurately reconstructed: direction, sense and position in the detector volume.
In order to optimize the track reconstruction and to fully exploit the data from the MIMAC detector, we developed a likelihood method dedicated to the track reconstruction.
This likelihood approach requires a full simulation of track measurements with MIMAC in order to compare real tracks to simulated ones. Finally, we found that the MIMAC
detector should have the required performance to perform a competitive directional detection of Dark Matter.
\end{abstract}
\maketitle
%%-----------------------------
%%      your text
%%-----------------------------

\section{Introduction}

Directional detection is a promising search strategy to discover galactic Dark Matter (\cite{spergel}).  
Taking advantage on the rotation of the Solar system around the Galactic center through the Dark Matter halo, it allows to show a direction 
dependence of WIMP events (\cite{billard.disco}). To do so, one has to be able to measure the energy and the three dimensional tracks of the recoiling nuclei at low energy $\sim O(10)$ keV. Indeed,
 like in  the case of direction-insensitive detection experiment, most of the WIMP events lies at low energy suggesting the need for a low energy threshold.
  However, in the case of a Fluorine
target at 50 mbar, the mean track length of a 10 keV recoil is of the order of 300 $\mu m$ and about 3 mm at 100 keV. Then, directional detection of Dark Matter
requires very sensitive experiment combined with highly performant technology. In this context, the MIMAC project has been proposed (\cite{mimac}) in order to achieve this
goal. In this paper, the track reconstruction strategy and the expected performance of the MIMAC detector are presented.
 
\section{MIMAC prototype and 3D track measurement}

The MIMAC prototype is the elementary chamber of the future large matrix. It allows the possibility to show the ionization and track measurement performance needed to
 achieve the directional detection strategy.
The primary electron-ion pairs produced by a nuclear recoil in one chamber of the matrix are detected by drifting the electrons to the grid of a bulk micromegas (\cite{bulk})
 and producing the avalanche in a very thin gap (128 or 256$\mu$m). 
 The electrons move towards the grid in the drift space and are projected on the pixelized anode thus allowing to get 
information on the X and Y coordinates.
To access the X and Y dimensions, a bulk micromegas (\cite{paco}) with a 10 by 10 cm$^2$ active area, segmented in pixels with an effective pitch of 424 $\mu$m
 is used as 2D readout.
 In order to reconstruct the third dimension Z of the recoil, a self-triggered electronics has been developed. It allows
  to perform the anode sampling at a frequency of 50 MHz.
This includes a dedicated 16 channels ASIC (\cite{richer}) associated to a DAQ (\cite{bourrion}). 

The ionization energy measurement is done by using a charge integrator connected to the grid wich is sampled at a frequency of 50 MHz. Then, to recover kinetic
energy of the recoiling nucleus, one has to know accurately the value of the Ionization Quenching Factor (IQF) (\cite{guillaudin}).

To summarize, the MIMAC readout system is able to provide a large number of observable which are the position of the center of gravity in the (X,Y) plane and the width along
the X and Y axis for each time slice. Moreover, if the charge integrator response is perfectly known, one should be able to recover the deposited charge in each time slice.
This leads us to the conclusion that the MIMAC readout provide us with a number of observables $\rm N_{obs}$ which grows with the number of time slice $\rm N_{slice}$ as: 
$\rm N_{obs} = 1 + 5\times N_{slice}$, with 1 refering to the observable $\rm N_{slice}$ itself.

 \begin{figure}[t]
\begin{center}
\includegraphics[scale=0.62,angle=0]{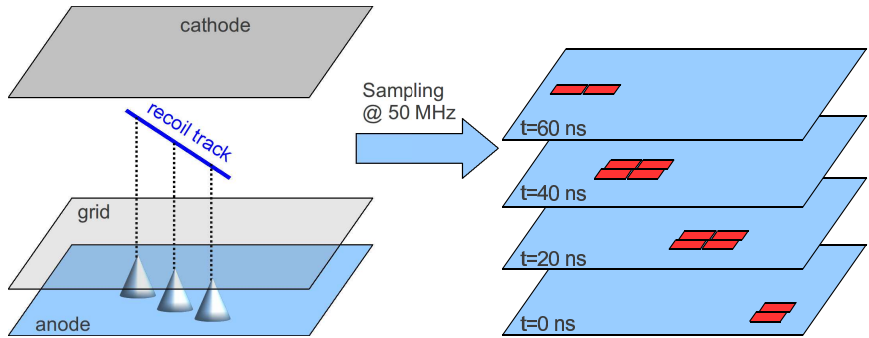}
\caption{Track reconstruction in MIMAC. The anode is scanned every 20 ns and the 3D track is recontructed,
 from the evolution in time of the collected charges on the (X,Y) plan, providing the drift velocity of the electrons is known.}  
\label{fig:MIMACDAQ}
\end{center}
\end{figure}

\section{Track measurement systematics}

\begin{figure}[t]
\begin{center}
\includegraphics[scale=0.30,angle=0]{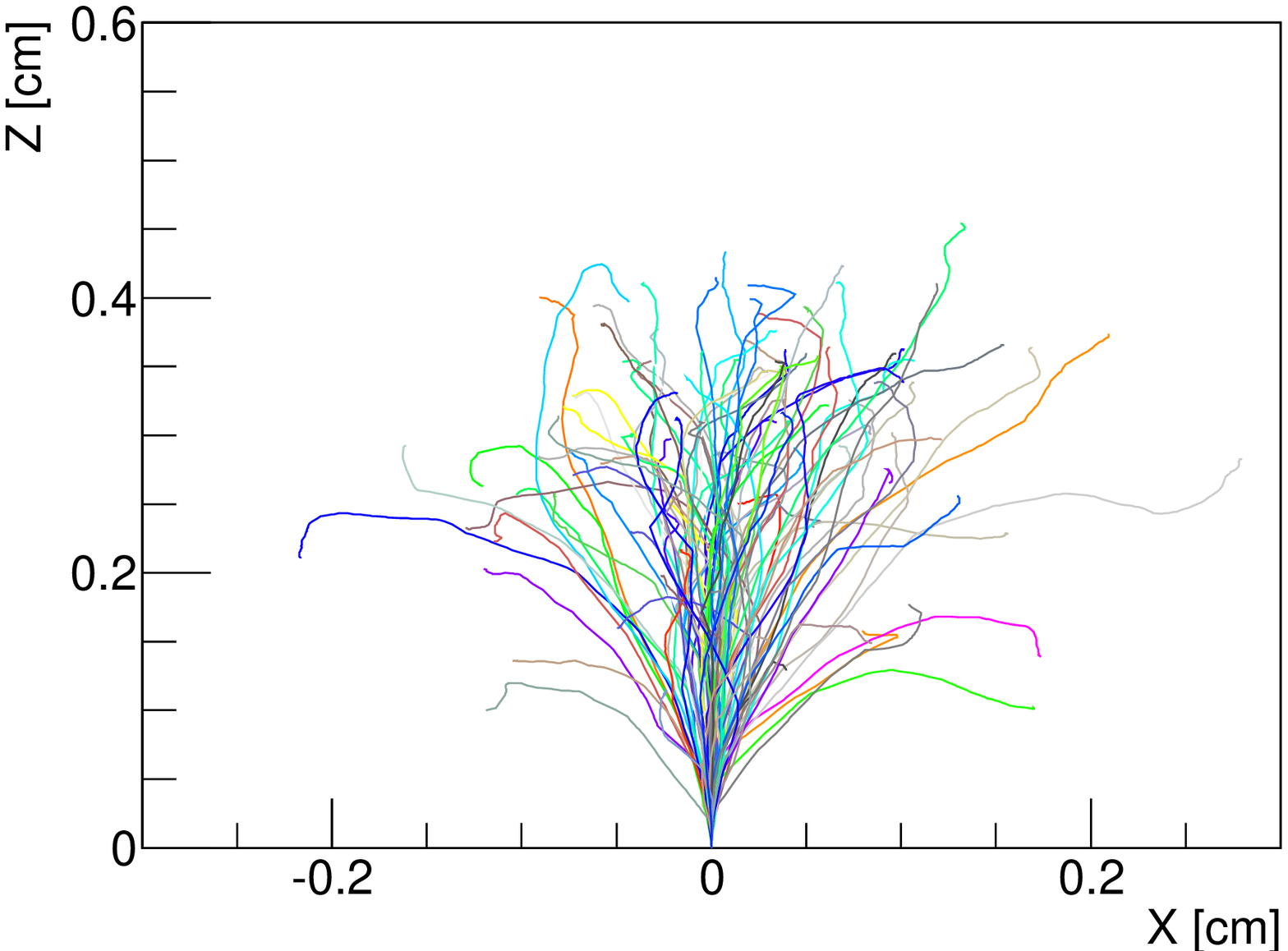}
\includegraphics[scale=0.6,angle=0]{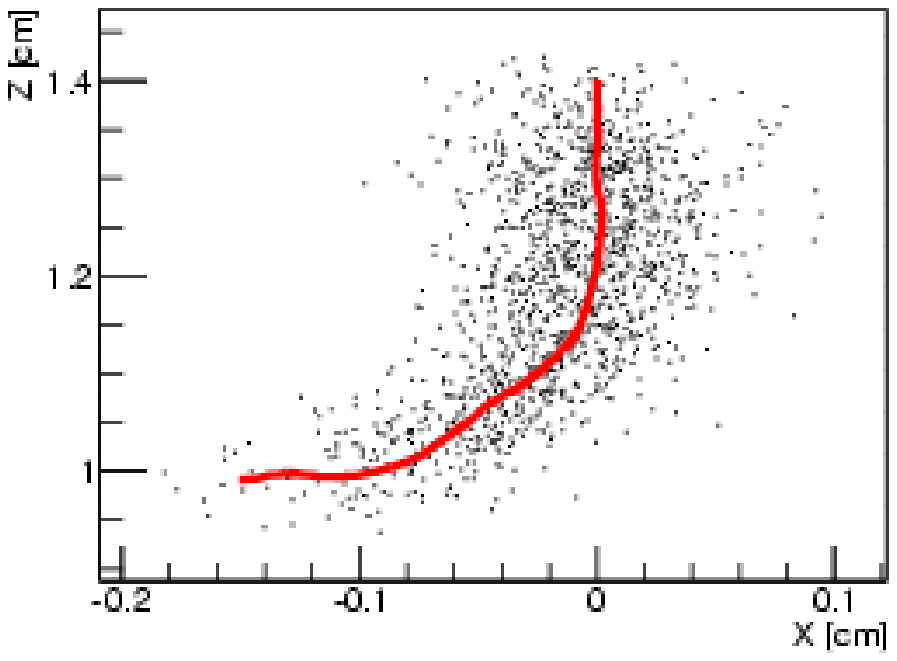}
\caption{Left: Representation in the (X,Z) plane of 100 simulated tracks of a Fluorine recoil with a kinetic energy of 100 keV in a gaz mixture of 
70\%CF$_4$ + 30\%CHF$_{3}$ at 50 mbar. Right: Representation in co-mobile coordinates of a single recoil track (red solid line) with its primary electron cloud (black dots).}  
\label{fig:Tracks}
\end{center}
\end{figure}

The performance of directional detection, in terms of resolutions, are obviously strongly correlated to the track properties of the target nucleus and to 
the tracking performance of the detector. 
In this section, we give a small review of some physical processes which contribute to the systematics associated to a track measurement in a TPC and more precisely
in the context of the MIMAC detector. Unless otherwise stated, we consider a gaz mixture composed of 70\% of CF$_4$ and 30\% of CHF$_3$ at 50 mbar.\\
On the left panel of figure \ref{fig:Tracks}, we have plotted 100 simulated tracks, using the SRIM software (\cite{srim}), corresponding to a Fluorine recoil at 100 keV. As it can be seen,
the angular dispersion (straggling), due to collisions with the other atoms of the gas, is very important and will strongly constrain the minimal angular resolution that a directional
detector could reach. Indeed, using a simple linear regression of these tracks, we found an intrisic angular dispersion of 18$^{\circ}$ at 100 keV and 25$^{\circ}$ at 10 keV.
Moreover, this large straggling also induces a intrisic limitation on the electronic/nuclear recoil discrimination based on the track length versus Energy
discrimination.\\
Another source of systematic is the number of primary electron generated along the track of the recoiling nucleus. Indeed, the number of primary electrons will impact both
the energy resolution and the track spatial resolution. However, the main contribution to the energy resolution 
 is the spread of the ionization quenching
factor, which is about $30\%$ for a Fluorine recoil at 20 keV, according to SRIM simulations.\\
It is compulsory to correctly evaluate the drift velocity of electrons and the diffusion coefficients to reconstruct the track especially if one tries to recover the Z
position of the track using the spread of electrons on the (X,Y) plane. In our case, according to Magboltz (\cite{magboltz}) calculations of the transverse diffusion coefficient is about 250 $\rm \mu m/\sqrt{cm}$ and leads to a spread
of the electron of about 1 mm after a drift distance of 16 cm. As an illustration of the effect of electron diffusion, we have shown on the right panel of figure 
\ref{fig:Tracks} the representation, in co-mobile coordinates, of a single Fluorine recoil track (red solid line) with the diffused primary electrons (black dots). It can be
noticed that the electron diffusion will smooth the recoil track and make the detector insensitive to the small deflections of the track.

Then, the track simulation software that we have developed is based on SRIM (\cite{srim}) for the track simulation, Magboltz (\cite{magboltz}) for an accurate estimation of the drift velocity and
dispersions, and on a software simulating the readout of the MIMAC detector.

\section{Likelihood approach for track reconstruction}

 \begin{figure}[t]
\begin{center}
\includegraphics[scale=0.60,angle=0]{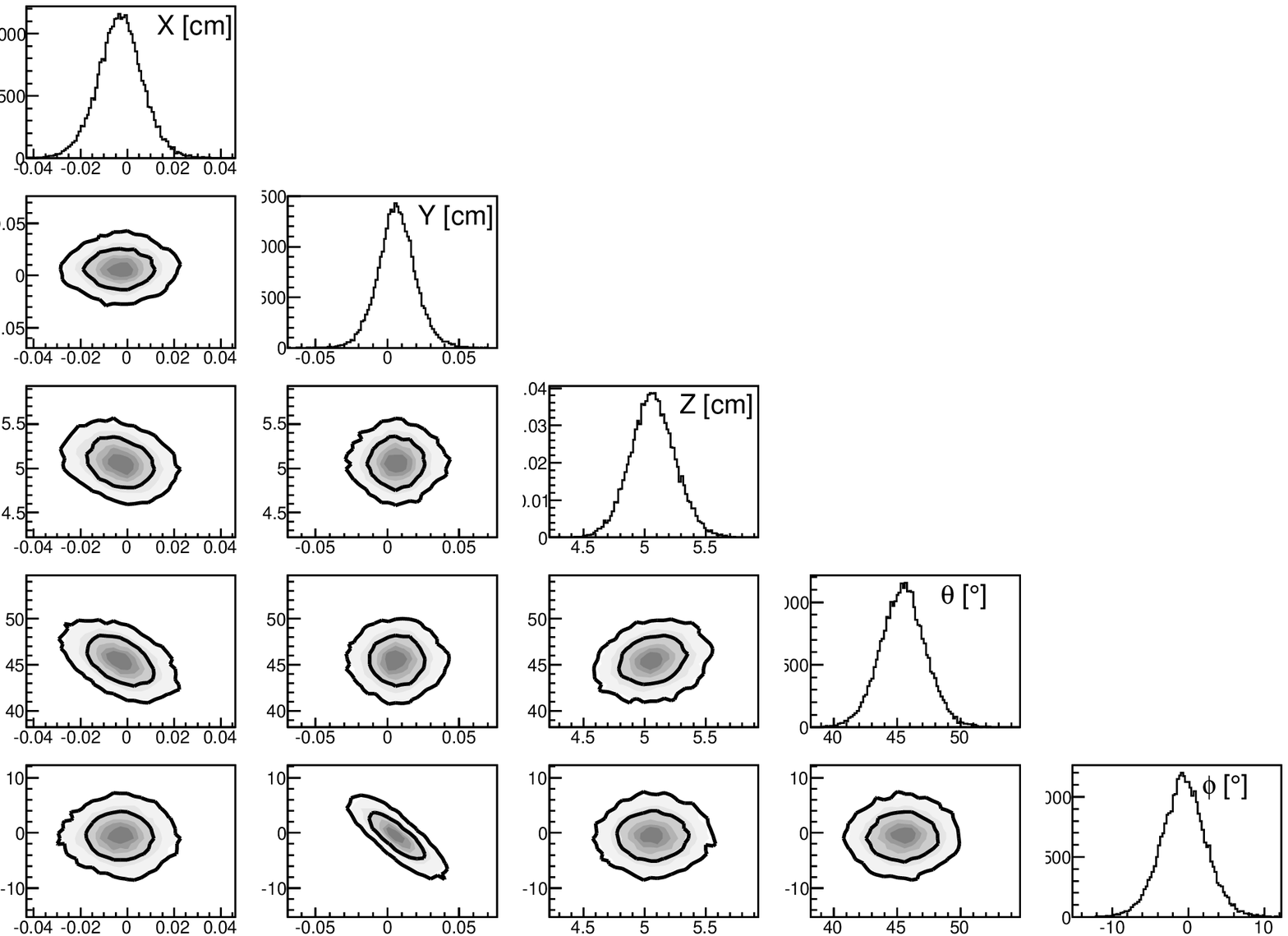}
\caption{Marginalized distributions (diagonal) ad 2D correlations (off-diagonal) plots of the likelihood function associated to a proton recoil of 100 keV generated at
the position \{X=0,Y=0,Z=5cm\}, and in the direction $(\theta= 45^{\circ},\phi = 0^{\circ})$ in a pure C$_4$H$_{10}$ gaz at 50 mbar.
 The likelihood has been sampled using a Markov Chain Monte Carlo algorithm (\cite{billard.ident}).}  
\label{fig:MCMCHydrogen}
\end{center}
\end{figure}

A straightforward track reconstruction algorithm consists of using a three dimensional linear fit of the center of gravity of each time slice. However, we found that this method
failed at recovering the correct $\theta$ angle due to the longitudinal diffusion of the electrons. In order to recover correctly the track properties, two strategies can be
used:
\begin{itemize}
\item If one is able to get the Z position of the track in the detector, {\it e.g.} by using a PMT sensitive to the primary scintillation, the electron diffusion can be
correctly estimated and substracted to the measured one
\item If no additional readout is used, the only way to recover without bias the track properties, is to fit all the parameters at the same time, using a likelihood approach.
This is the strategy that we have developed and which is presented hereafter.
\end{itemize}

As shown in the previous section, we can simulate any kind of recoiling track according to the following input parameters: X, Y, Z, $\theta$, $\phi$ and the sense ${S}$
 (upward or downward). Then, the main idea of this likelihood approach is to compare the measured track to the simulated ones.
The analysis is based on the computation of the likelihood function
 $\mathscr{L}(X, Y, Z, \theta, \phi| {S})$ defined as,
 \begin{equation}
\mathscr{L}(X, Y, Z, \theta, \phi| {S}) = P(N_{slice})\prod_{n=1}^{N_{slice}}P(X_n^{bary})P(Y_n^{bary})P(\Delta X_n)P(\Delta Y_n)P(Q_n)
 \end{equation}
where the different probability $P$ are estimated using a chi-square function defined in the case of the $\Delta X_1$ observable, for example, as:
\begin{equation}
P(\Delta X_1) = \exp\left\{-\frac{1}{2}\left(\frac{\Delta X_1 - \hat{\Delta X_1}}{\hat{\sigma}_{\Delta X_1}}\right) \right\}
\end{equation}
 where $\hat{\Delta X_1}$ and $\hat{\sigma}_{\Delta X_1}$ correspond to the expected mean value and RMS of $\Delta X_1$ according to the given set of parameters. Then, 
this method requires a large number of simulated tracks to get an accurate estimation of the expected mean value and RMS associated to each observable.
 Also, it assumes a
Gaussian distribution of the different observable which appears to be the case except for tracks very close to the anode, {\it i.e.} $\lesssim$ 1 cm.

As an illustration of the method, we show on figure \ref{fig:MCMCHydrogen} the likelihood function, sampled using a Markov Chain Monte Carlo (MCMC) algorithm,
associated to a proton recoil of 100 keV in a pure $\rm C_4H_{10}$ gas at 50 mbar. The latter was simulated at $\{X=0,Y=0,Z=5 \text{cm}\}$ in the
direction ($\theta= 45^{\circ}, \phi = 0^{\circ}$) and going downward. As one can see from figure \ref{fig:MCMCHydrogen}, the five parameters are strongly and consistently
constrained according to their input values. Even the $\theta$ angle and the $Z$ position which are the most challenging parameters are fully recovered. This study, even if
it was applied on a proton track which is not of interest in our case, highlights the fact that a likelihood approach to track reconstruction of a few tens of keV allows to
recover consistently the track properties. Another interest of using such a likelihood analysis is that for each reconstructed track, one can get both the best fit value and
 the error bars on each parameters, taking into account all the systematics associated to the track detection.

\section{Expected resolutions}

In order to estimate the expected performance of the detector, in terms of spatial and angular resolution, systematical studies have been done. To do so, we have simulated
about one thousand tracks generated isotropically in the lower half sphere (going downward) at a fixed Z coordinate and energy.
Then, for each track, we used a maximization algorithm in order to retrieve the
maximum likelihood location in the five dimensional parameter space. Then, the spatial resolutions $\sigma_{x,y,z}$ are simply obtained using the distribution 
$f(X_i) = X_i - X^0_i$ where $X^0_i$ correspond to the input value of each single track and \{i = x,y,z\}. In the case of a Fluorine recoil in our gas mixture 
(70\% CF$_4$ +
30\%CHF$_3$), we found that $\sigma_{x,y}$ is about 1 mm at 20 keV and 0.4 mm at 100 keV and that it is independent from the Z coordinate. About $\sigma_z$, we found a
resolution of about 1.5 mm at 40 keV and 1 mm at 100 keV, with a weak dependence on the Z coordinates.
 The fact that the spatial resolutions
are in the order of the millimeter will allow us to perform an accurate three dimensional fiducialization of the detector volume to prevent from surface events.

\begin{figure}[t]
\begin{center}
\includegraphics[scale=0.30,angle=0]{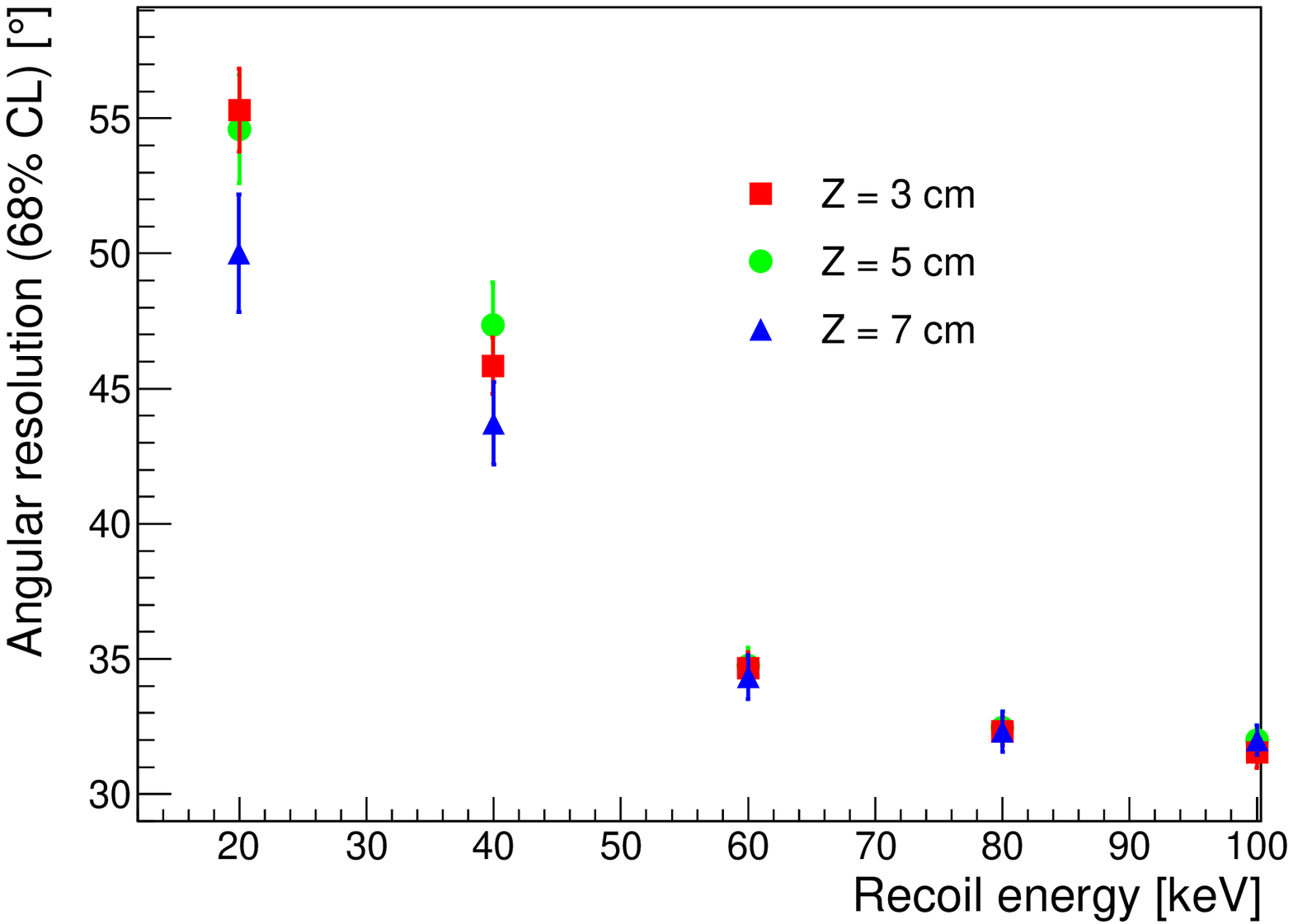}
\includegraphics[scale=0.30,angle=0]{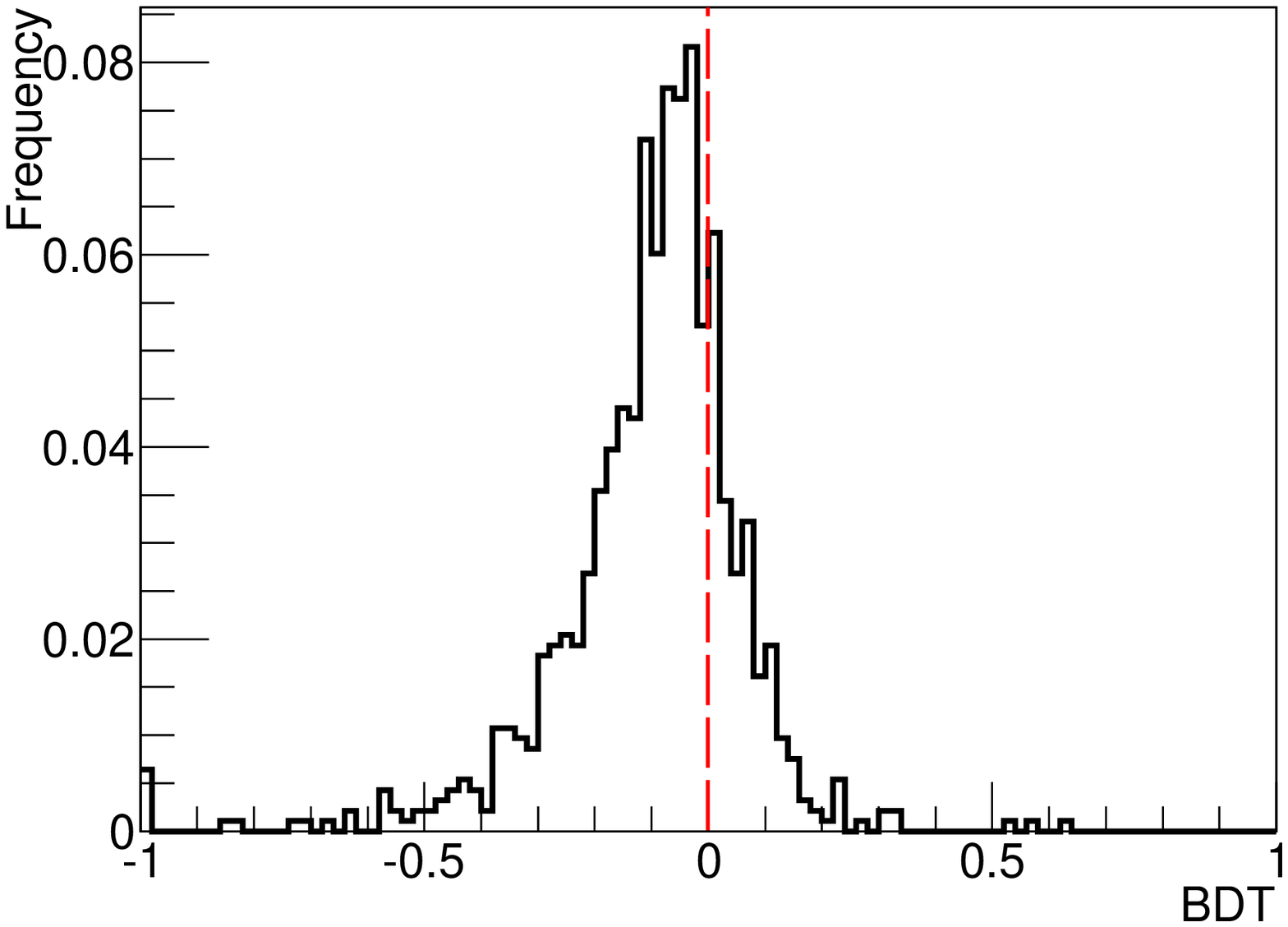}
\caption{Left: Angular resolution $\sigma_{\gamma}$ as a function of the fluorine recoil energy for three different values of Z: Z=3 cm (red squares), Z=5 cm (green dots) and
Z=7 cm (blue triangles). Right: Distribution of the output BDT value after each track reconstruction. The BDT distributions corresponds to the case of a fluorine recoil at
100 keV at Z = 1.3cm, in the direction ($\theta= 45^{\circ}, \phi = 0^{\circ}$) and going downward.}  
\label{fig:Resolutions}
\end{center}
\end{figure}

The angular resolution is estimated with the construction of the distribution $f(\gamma)$ where $\gamma$ is the angle between the true direction of the track and its reconstructed
direction. In the case of a Gaussian resolution, which turned out to be the case in our study, the distribution $f(\gamma)$ is described by the following expression:
\begin{equation}
f(\gamma) \propto \sin\gamma\exp\left\{-\frac{1}{2}\left(\frac{\gamma}{\Delta \gamma}  \right)^2 \right\}.
\end{equation}
The angular resolution $\sigma_{\gamma}$ at 68\% C.L. is then obtained by solving,
\begin{equation}
\int_{0}^{\sigma_{\gamma}}f({\gamma})d\gamma = 68\%
\end{equation}
The result is shown on the left panel of figure \ref{fig:Resolutions}. One can see that $\sigma_{\gamma}$ depends strongly on the energy of the recoiling nucleus. Indeed,
the angular resolution is about 55$^{\circ}$ at 20 keV while it is about 30$^{\circ}$ at 100 keV. However, one can see that it does not depend on the Z coordinate of the
track in the 3 cm to 7 cm range. As we have an analytical expression of the distribution $f(\gamma)$, we used the frequentist confidence belt method to calculate the error
bars associated to each estimate of the angular resolution.

\section{Toward a sense recognition?}

The sense recognition capability is considered as the most challenging experimental issue of directional detection. Indeed, to reach this ability, detectors have to be very sensitive to
any sources of asymmetry between the two hypothesis: is the track going downward or upward ($S$ = Down or Up)? The reason why experimentalist believe that it may be possible to achieve sense
recognition is that there is, indeed, two sources of asymmetry between the beginning and the end of the track:
\begin{itemize}
\item Shape asymmetry: As it is shown on the left and right panels of figure \ref{fig:Tracks}, more deflections are expected at the end of the track than at its
beginning. Of course, due to electron diffusion and the pitch size of the detector readout, this effect might be strongly smeared out. However, even if a realistic detector won't
be able to measure all the track deflections, it should be at least sensitive to an asymmetry in the spread of the track between the begining and the end of the track.
\item Charge asymmetry: At low energy, $\lesssim O(1)$ MeV, the ionization dE/dx is expected to be higher at the begining of the track rather than at the end. This should
lead to an asymmetry in the charge integration readout which is a function of the projection of the collected charges along the Z axis. Obviously, if the track is parallel to
the anode, the detector won't be sensitive to this asymmetry.
\end{itemize}

To optimize the sense recognition efficiency, we used a high dimensional multivariate analysis, namely a Boosted Decision Tree (BDT) (\cite{tmva}). It
 can be seen as a classifier signal/background or in our case Up/Down. Its principle is based on the optimisation of linear cuts on the different observables which are taken
 into account in the analysis. A considerable interest with the BDT, in comparison with a neural network for example, is that adding a new observable, even if its discrimination power
 is very weak, will never degrade the BDT efficiency and won't change the calculation time. That way, Boosted Decision Tree can easily handle a large number of
 observables and requires a very little calculation time. However, like any multivariate analysis algorithm, one has to check for overtraining. This check can be done
 by using a Kolmogorov test statistic by comparing the distributions of the BDT value obtained on the ``training'' and the ``test'' samples.
 
 In our case, we used the Boosted Decision Tree as a sense recognition discriminant by first finding the maximum Likelihood according to the two hypothesis: Up and Down.
 Then, we generate a large sample of tracks (around 10,000) corresponding to the best fit of the two hypothesis. One half of the simulated tracks are used to train the Boosted Decision
 Tree algorithm and the other half to check for overtraining. Finally, the final result of the Boosted Decision Tree
  is a function defined as: ${\rm BDT} =  f(\vec{X})$, where BDT represents the discriminating variable and $\vec{X}$ all the used observables. That way, if the track is
  going upward or downward, the BDT value should be positive or negative respectively.
  
  Finally, by computing ${\rm BDT} =  f(\vec{X})$ to the real or simulated measured track, one can estimate if the latter is going downward or upward. As a working example, we
  have shown on the right panel of figure \ref{fig:Resolutions} the BDT distribution associated to 1000 simulated tracks generated at 100 keV, in the direction $(\theta =
  45^{\circ},\phi=0^{\circ})$, going downward and at 1.3 cm of the anode. As one can see from the right panel of figure \ref{fig:Resolutions}, 77\% of the tracks have a negative
  BDT value, thus corresponding to a 77\% sense recognition efficiency for this given set of parameters (position, direction and energy).

\section{Conclusion}

In this paper, we have shown that, in the case of the MIMAC detector with the pixelized anode and the charge integrator readout, a likelihood approach dedicated to the track
reconstruction is very well suited to retrieve the track properties. Indeed, with a track simulation software using SRIM, Magboltz and the MIMAC DAQ simulation we are able to both recover the track
properties without bias and to estimate the error bars associated to each parameter of interest, taking into account all the systematics of a nuclear recoil track detection
with the MIMAC detector. That way, we found that the spatial resolutions are of the order of 1 mm opening the possibility to use an accurate three dimensional fiducialization of the detector volume.
 Also, the angular resolution was estimated to be between 55$^{\circ}$ and
 30$^{\circ}$ for recoiling energies from 20 keV to 100 keV respectively. Finally, we have also shown the possibility to perform a sense recognition of the recoiling tracks.
 Indeed, with the use of a large number of observables ($\rm N_{obs}$) within the framework of Boosted Decision Tree method, we found 77\% sense
 recognition efficiency with a 100 keV Fluorine recoil. To conclude, the MIMAC detector combined with a likelihood track reconstruction method should be able to perform
 a highly competitive directional detection of Dark Matter.

%%-----------------------------
%%      your bibliography
%%-----------------------------

\end{document}